\documentclass[aps,pre,twocolumn,"Cornell University, Ithaca, NY, 14853"]{revtex4}
\usepackage{amsmath}
\usepackage{empheq}
\usepackage[parfill]{parskip}
\usepackage{graphicx}
\usepackage{caption}
\usepackage{color}
\usepackage{soul}
\usepackage{etoolbox}

\begin{document}
\newcommand{\bi}{\begin{itemize}}
\newcommand{\ei}{\end{itemize}}
\newcommand{\be}{\begin{equation}}
\newcommand{\fe}{\end{equation}}
\newcommand{\noi}{\noindent}
\newcommand{\co}[1]{(\textbf{#1})}

\setlength\parindent{20pt}

%Title of paper
\title{Transient dynamics of pulse-coupled oscillators with nonlinear charging curves}

\author{Kevin P.  O'Keeffe}
\affiliation{Department of Physics, Cornell University, Ithaca, NY 14853, USA}

\date{\today}

\begin{abstract}
We consider the transient behavior of globally coupled systems of identical pulse coupled oscillators. Synchrony develops through an aggregation phenomenon, with clusters of synchronized oscillators forming and growing larger in time. Previous work derived expressions for these time dependent clusters, when each oscillator obeyed a linear charging curve. We generalize these results to  cases where the charging curves have nonlinearities.

\end{abstract}

\date{\today}
\pacs{05.45.Xt, 05.70.Ln}

%%%%%%%%%%%%%%%%%%%%%%%%%%%%%%%%%%%%%%%%%%%%%%%%%%%%%%%%%%%%%%%%

\maketitle

\section{Introduction}

During each heartbeat, thousands of pacemaker cells discharge in concert. This collective firing causes the contraction of cardiac muscles, which pump blood around the body. Should these firing fall out of step, heartbeats can become erratic, which inhibits blood flow.  In order to maintain healthy heart function, the pacemaker cells must maintain their synchronous firing.  

In 1975, Peskin gave the first mathematical analysis of the pacemaker as a self-synchronizing system \cite{peskin}. He modeled the pacemaker cells as leaky 'integrate-and-fire' oscillators that communicate with each other by firing sudden impulses. He then conjectured that a population of identical leaky oscillators with all-to-all pulsatile coupling would self-organize into synchrony for all $N \ge 2$ and for almost all initial conditions.  Mirollo and Strogatz \cite{strogatz_and_mirollo} later proved this conjecture.

Since then, pulse-coupled oscillators have been used as models in many other contexts, for example, sensor networks \cite{sensor_networks_time}, low-powered radio transmission \cite{low_powered_radios}, firing neurons \cite{neurons, ermentrout_book}, earthquakes \cite{earthquakes}, and economic booms and busts \cite{economics}. For greater realism, the associated theoretical work relaxed Peskin's original assumptions, allowing for local coupling in lattices or networks, non-instantaneous interactions, and so on \cite{temporal_delays, delays,ring1, ring2, SOC, SOC1,coexistence, async_state, unstable_attractors, networks, arenas_networks, ref1, ref2, ref3, ref4, ref5, ref6, ref7,ref8, ref9}. 

Yet even within the simplified context of Peskin's all-to-all model, unanswered theoretical questions remain.  In particular, little is known about transient dynamics: in a self-synchronizing system, what does the buildup to synchrony look like? A first step in this direction was presented in \cite{sync_as_aggregation}. It was shown that synchrony developed through clustering; oscillators start to synchronize in small groups, which grow steadily larger over time. Using tools from aggregation theory \cite{krapivsky_book}, this clustering was described quantitatively. In the analysis, it was  assumed that each oscillator had a linear charging curve. This idealization is appropriate for electronic oscillators such as those in sensor networks, but not for biological oscillators, like the aforementioned cardiac pacemaker cells or firing neurons. We here extend the analysis in \cite{sync_as_aggregation} to explore the manner in which these more complicated oscillators achieve synchrony. \\

%%%%%%%%%%%%%%%%%%%%%%%%%%%%%%%%%%%%%%%%%%%%%%%%%%%%%%%%%%%%%%%%

\section{The model} 
We consider $N \gg 1$ identical oscillators coupled all-to-all. Each oscillator is characterized by a voltage-like state variable $x_i$, which increases from a baseline value of $0$ to a threshold set to 1, according to $\dot{x_i} = S_0 - \gamma x_i$. When an oscillator reaches threshold it does two things: (i) It fires a pulse of size $1/N$. This pulse is received by all other oscillators instantaneously, causing them to discontinuously raise their voltage from $x_j$ to $\min(x_j + 1/N, 1)$. This way, oscillators never exceed the threshold value of $1$. To avoid complications with chain reactions of firing oscillators, we assume any oscillators which reach threshold by receiving a pulse, do not themselves fire until the \textit{next} time they reach threshold. (ii) The firing oscillator then resets its voltage to $0$, along with any secondary oscillators that were brought to threshold. These oscillators then begin their next cycle synchronized.

If $j > 1$ oscillators reach threshold together, each one fires, so that the pulse has total size $j/N$ (although we later consider other types of pulse).

We note that there is some parameter redundancy, since by rescaling time we could set $S_0 = 1$ without loss of generality. For reasons that will become clear later, a different choice of $S_0$ is more convenient, so we leave it as a free parameter for now. We remark however that $S_0$ must be chosen so that $\dot{x_i} > 0$ for $0 \leq x \leq1$.

%%%%%%%%%%%%%%%%%%%%%%%%%%%%%%%%%%%%%%%%%%%%%%%%%%%%%%%%%%%%%%%%

\section{Results}

%%%%%%%   Defintions      %%%%%%%

Assume the initial voltages of the oscillators are drawn uniformly at random. How will the dynamics unfold? At the beginning, the oscillators simply increase their voltage according to $\dot{x_i} = S_0 - \gamma x_i$. Then the first oscillator reaches threshold, fires a pulse, and perhaps brings some other oscillators to threshold. As described, these oscillators begin their next cycle in step. The primary, firing oscillator, and the secondary oscillators it incited to threshold, form a synchronous \textit{cluster}. 

As time goes on, other oscillators start firing pulses and start absorbing oscillators which are close enough to threshold. More clusters of synchronized oscillators emerge. In turn, these clusters start reaching threshold and absorbing \textit{other} clusters, growing progressively larger. We note that clusters can only ever increase in size. They can never break apart because (a) the oscillators are identical, and therefore sync'd oscillators have the same speed, and (b) all oscillators receive the same number of  pulses (thanks to the global coupling).

The picture is now clear; the system synchronizes through an aggregation phenomenon. Clusters of sync'd oscillators form and get steadily bigger by coalescing with each other. At any time $t$ therefore, there are clusters of various sizes. Let $N_j(t)$ denote the number of clusters of size $j$ at time $t$:  $N_1$ is the number of singletons, $N_2$ is the numbers doublets, and so on. These $N_j$ are correlated random quantities. They are correlated because oscillators belonging to clusters of one size are unavailable to clusters of another size, and they are random because of the initial conditions. 

To analyze the system's dynamics, we imagine $N_j \gg 1$ so that fluctuations from different realizations of the system are small. Of course, this condition cannot be satisfied for every $j$, at all $t$. For example, at the final stages of the process, there will be a small number of very large clusters. We therefore restrict our attention to the portion of the process where $N_j \gg 1$ is  approximately true -- the opening and middle game, as opposed to the end game.

But how does the end game play out? That is, how does the process terminate? Strogatz and Mirollo \cite{strogatz_and_mirollo} showed that for $\gamma > 0$ and pulse size $> 1/N$, then full sync is guaranteed for all IC except for a set of measure zero; the clustering continues until there is one giant cluster of size $N$. For other values of $\gamma$ and other pulse sizes, full sync is possible, but not certain to occur.

In this work, we focus only on the transient dynamics, the evolution \textit{to} synchrony. So from now on we implicitly assume we in the early and middle stages of the process, where $N_j \gg 1$ is a valid approximation. We then use ensemble averages to define the \textit{individual cluster densities},

\be
c_j := \langle N_j \rangle / N.
\fe

\noindent
We then make the following strong assumptions: (i) fluctuations about the ensemble averages are small, $N^{-1} N_j = c_j + O(N^{-1/2})$, and that (ii) different cluster densities are asymptotically uncorrelated, $N^{-2} N_i N_j = c_i c_j + O(N^{-1/2})$.

We can use these $c_j$ to define a disorder parameter for our system. This is the \textit{total cluster density},
 
 \be
 c = \sum_j c_j.
 \fe
 
\noindent 
It measures of the total fragmentation of the system, which we interpret as a kind of disorder. To see this, consider that at $t=0$, there are $N$ singletons, so $c_1 = 1$, and $c_j = 0, \;  \forall j \ne 1$. This means that $c(0) = 1$, correctly identifying that the system begins maximally disordered. At the other extreme as $t \rightarrow \infty$, we know there is one giant cluster of size $N$, so $c = 1/N \approx 0$ for large $N$. Hence $c$ decreases from $1$ to $0$ as the system evolves from complete disorder to full synchrony.

\subsection{Total Cluster Density}

%%%%%%%   Rate eqn for c      %%%%%%%

We first analyse $c$. It obeys the following rate equation, where $R_i$ is the rate at which clusters of size $i$ fire, and $L_i$ is the number of clusters absorbed during such a firing, for $i = 1, \dots , N$:

 \be
 \dot{c} = - \sum_i R_i(t) L_i(t)
 \label{c_rate_eqn}
 \fe

 %%%%%%%   Find L_i      %%%%%%%

\noindent 
To find $L_i(t)$, we first define the 'voltage-density' $\rho_j(x,t)dx$ to be the number of $j$-clusters with voltage between $x$ and $x+dx$ at time $t$. This has the normalization condition $\int_{0}^1 \rho_j(x,t) dx = N_j$. Now, when an $i$-cluster fires, all clusters on the interval $[1-i/N, 1)$ will be absorbed. This means,

\be
 L_i(t) = \sum_j  \int_{1-i/N}^{1} \rho_j(x,t) dx.
\fe

\noindent
We now make an approximation. As stated earlier, we are only interested in transient time scales  -- the opening and middle game. In this regime, most clusters will be small relative to the system size: $j \ll N$. This lets us approximate the integral above, $\int_{1-i/N}^{1} \rho_j(x,t) dx \approx (i/N) \rho_j(x=1, t)$. Of course, this approximation will get worse as time goes on. We discuss this further in Section \ref{approximations}. Our expression for $L_i$ is then

\be
 L_i(t) = \frac{i}{N} \sum_j  \rho_j(x =1,t)
 \label{L_i_soln}
\fe

%%%%%%%   Find R_i      %%%%%%%

\noindent
To continue the analysis, we need to find $\rho_j(x,t)$. Its behavior is however complicated so we defer its calculation, and instead find the firing rate $R_i$. Naively, one might think that this is simply the flux of $i$-clusters at threshold: $N^{-1}(\rho_i v) |_{x=1}$ (where $N^{-1}$ is required, since $R_i$ measure the rate of firing of $c_i$, not $N_i$). However not every cluster that reaches threshold gets the chance to fire, since some will be absorbed. To account for this effect, we decompose the rate into 

\be
R_i = R_i^0 - R_i^a. 
\label{R_i_def}
\fe

\noindent
The term $R_i^0$ is a 'background' firing rate, where we pretend all oscillators get to fire even if they are absorbed. $R_i^a$ is the rate at which $i$-clusters are \textit{being} absorbed by other clusters of various sizes, and hence deprived of their chance to fire. 

We start with $R_i^0$. To be clear, by background firing rate, we mean the rate $i$-clusters would fire at, if \textit{every} oscillator fired a pulse when it reached threshold. That is, imagine relaxing our imposition that any secondary oscillators that reach threshold by virtue of a pulse do not fire. In that case, 

\be
R_{i}^0 = N^{-1}(\rho_i v) \big|_{x=1}.
\fe

\noindent
The speed $v$ of each cluster is non-trivial. This is because in addition to its natural speed $v_0 = \dot{x} = S_0 - \gamma x$, each oscillator receives a steady stream of pulses from firing clusters which increase its voltage:

\begin{equation}
v(x,t) = v_0(x) + v_{pulse}(t).
\label{v_def}
\end{equation}

\noindent
This "pulse velocity" due to the firing of just $j$-clusters will be  (absolute number of pulses per sec) $\times$ (distance per pulse). Since $R_j$ is the firing rate of $c_j$, $R_jN$ is the absolute number of pulses, while the distance per pulse is $j/N$. To find the total pulse speed we then sum over all $j$-clusters: $ \sum_j (N R_j)(j/N)$, giving 

\begin{equation}
v_{pulse}(t) = \sum_j j R_j(t).
\end{equation}

\noindent
Our next target is the absorption rate $R_i^a$. The calculation is similar to finding $L_i$, and is given by $R_i^a = \sum_j R_j \int_{1-j/N}^1 \rho_i(x,t) dx$, which after approximating the integral as before gives,

\be
R_i^a = \sum_j R_j (j/N) \rho_i(x=1,t).
\fe

\noindent
Substituting $R_i^0$ and $R_i^a$ into \eqref{R_i_def} finally gives

\be
R_i = \frac{S_0-\gamma}{N} \rho_i(x=1, t).
\label{R_i_soln}
\fe

 %%%%%%%   Find rho_j(x,t)     %%%%%%%
 
\noindent
We now analyze $\rho_j(x,t)$. In principle, it satisfies the the continuity equation with appropriate terms for the absorption of $j$-clusters at threshold, and the formation of $j$-clusters from smaller clusters:

\begin{align}
   & \dot{\rho_j} + \partial_x(v \rho_j) + Absorption + Gain  = 0 \\
    &\partial_x(v \rho_j)|_{x=0} = \partial_x(v \rho_j) |_{x=1}
      \label{cont_eqn0}
  \end{align}

\noindent
 This PDE is difficult to solve generally, given the non-smooth behavior of oscillators when they reach threshold. To proceed, we make the problem simpler, by observing that the evolution of the system naturally divides into periods $\{ T_n \}$. We define a period to be the time take for the full population of oscillators to complete a voltage cycle. More carefully, $T_n$ is earliest time when every oscillator has completed $n$ cycles. 

We then solve the continuity equation during a given period, not worrying about what happens before or after. This lets us avoid the complication of the non-smoothness of the oscillators' behavior at the boundaries. We also neglect the absorption term. As previously discussed, when an $i$-cluster fires, only oscillators on $(1-i/N, 1]$ get absorbed. This is a small interval for the 'opening' and 'middle' game we are considering. Hence the absorption term is $0$ on most of $[0,1]$ and so we omit it.

But we still have to compute the gain term. This would involve enumerating all the ways a $j$-cluster can be formed, which, as we will later show, is combinatorially intensive. We can however neglect this cumbersome term entirely, by making the following key observation.

Looking at equations \eqref{L_i_soln}, \eqref{R_i_soln}, we see our desired quantities $R_j$ and $L_j$ depend only on the density of clusters at threshold: $\rho_j(x=1, t)$. Therefore, \textit{during each period}, $R_i$ and $L_i$ are only affected by $j$-clusters which existed \textit{at the start of that period}, which we call 'original' $j$-clusters. This is because any 'new' $j$-clusters won't reach threshold until the \textit{next} period. By 'new', we mean (a) $j$-clusters that fired during a period, didn't absorb any other clusters, and so returned to threshold, and (b) any $j$-clusters that were created by the firing and absorption of other smaller clusters. 

So for the purposes of calculating $\rho_j(x=1,t)$ \textit{during a given period}, there is a 'lightcone' between original and new $j$-clusters. We therefore need to  solve the continuity equation for the original $j$-clusters only, for which the gain term in zero. The problem is then given by \eqref{cont_eqn1} below, where $v(x,t)$ is given by \eqref{v_def}, $f_0(x)$ is the initial distribution of $\rho_j^{original}$, and the Heaviside functions $H(x), H(1-x)$ are included to confine the I.C. to the interval $[0,1]$.

%\begin{equation}
\begin{align}
    &\dot{\rho_j}^{original} + \partial_x(v \rho_j^{original}) = 0 \nonumber \\
    &\rho_j^{original}(x,0) = f_0(x) H(x) H(1-x)
      \label{cont_eqn1}
  \end{align}
%\end{equation}

\noindent
We don't yet know the speed $v(x,t)$. However its structure,  $v(x,t) = v_0(x) + v_{pulse}(t)$,  lets us derive an approximate solution for $\rho_j^{original}(x,t)$ given by \eqref{rho_period1} below. The derivation of this key result and the definition of $\Gamma(x,t)$ are shown in the Appendix.

\be
\rho_j^{original}(x,t) = e^{\gamma t} f_0 \Big( \Gamma(x,t) \Big) H \Big( \Gamma(x,t) \Big) H \Big( 1 -\Gamma(x,t) \Big).
\label{rho_period1}
\fe

\noindent
What all this means is, if we known $\rho_j^{original}(x,t)$ at the start of a period, then we know how it will evolve until that period ends. For later convenience, we introduce the following notation. Let $\tilde{x}$ denote that during a period, $x$ is held fixed at its value at the start of that period: $\tilde{x} = x(t = T_n)$ for $T_n < t < T_{n+1}$.

We next make the strong assumption that clusters of all sizes are distributed uniformly in voltage on $[0,1)$ at the start of each period: $\rho_j(x, t = T_n) = \tilde{N_j}$. Then in \eqref{rho_period1}, $f_0(x) = \tilde{N_j}$. We discuss the legitimacy of making this assumption in Section \ref{approximations}.

The Heaviside functions make \eqref{rho_period1} look complicated. But really, they only enforce that the $\rho_j^{original}$ is zero before and after the first and last $j$-cluster respectively.  We remark that as it stands, the solution \eqref{rho_period1} propagates into the unphysical $x \geq 1$ regime. But we of course restrict our attention to just $x \in [0,1]$.

\begin{figure}[h!]
  \centering
\includegraphics[width=8cm]{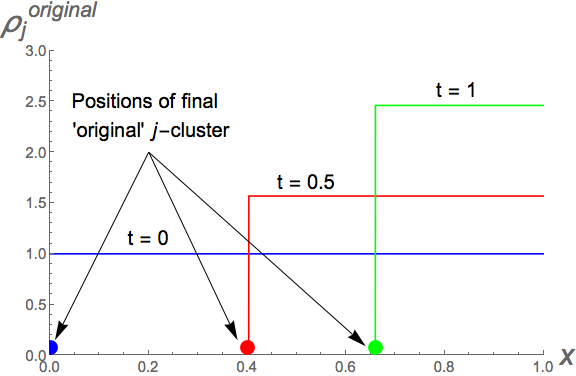}  
\caption{Evolution of voltage density of original $j$-clusters during a period, with initial condition $\rho_j^{original}(x,0) = 1$.}
\label{rho_evolution}
\end{figure}

The behavior of $\rho_j^{original}(x,t)$ during each period is therefore simple. The density at each point $x$ simply grows at rate $e^{\gamma t}$ until it drops discontinuously to $0$, as the final 'original' $j$-cluster passes by (by 'final' we mean the $j$-cluster which had the lowest voltage at the start of the period). This behavior is shown in Figure~\ref{rho_evolution}.

%%%%%%%%%%%    FIG 1      %%%%%%%

%%%%%%%   Find c     %%%%%%%

\noindent
Now that we have an expression for $\rho_j(x=1,t)$, which we have argued is $\rho_j^{original}(x=1,t)$, we can complete our expressions for $L_i$ and $R_i$ given by \eqref{L_i_soln} and \eqref{R_i_soln}. We then plug the results into \eqref{c_rate_eqn} to obtain our sought after rate equation for the disorder parameter $c(t)$,

\be
\dot{c} = -(S_0 - \gamma) e^{2 \gamma t} \tilde{c}.
\label{c_rate_eqn_final}
\fe

\noindent
which has solution,

\be
c(t) = \frac{\tilde{c}}{2 \gamma} \Big(  S_0 + \gamma + e^{2 \gamma t}(\gamma - S_0) \Big).
\label{c_soln1}
\fe

\noindent
We restate that equations \eqref{c_rate_eqn_final} and \eqref{c_soln1} are only valid during a given period. We can however use \eqref{c_soln1} to  find $c(t)$ for all $t$, by stitching solutions during successive periods together.

%%%%%%%   Find period     %%%%%%%

But we still don't know the periods $\{ T_n \}$ themselves. To find them, we need the speed $v$ as per \eqref{v_def}. Recalling $v_0 =  S_0 - \gamma x$, and substituting $R_i$ from \eqref{R_i_soln}, gives

\be
v(x,t) =(S_0- \gamma x) + (S_0-\gamma) e^{\gamma t}.
\fe

\noindent
We see that $v$ is the same during each period (i.e. there are no 'tilde' quantities, we denote different values during different periods.). This means that the length of each period is the same: $T_n = n T_0$. We can find $T_0$ from $T_0(S_0, \gamma) = \int_0^1 v(x) dx$. To compare the effects of different amounts of concavity on equal footing, we want $T_0 = 1$ for every $\gamma$. We can achieve this by selecting an appropriate value for $S_0$, which we have strategically left as a free parameter thus far. Doing the integral, this value for $S_0$ is

\be
S_0 = \frac{\left(e^{2 \gamma } + 2 e^{\gamma } -1\right) \gamma }{\left(e^{\gamma }-1\right) \left(e^{\gamma }+3\right)}.
\label{S_0}
\fe 

\noindent
We must be careful when using \eqref{S_0}. This is because for sufficiently negative $\gamma$, $S_0$ can become negative. While we have ensured the total speed $v = v_0 + v_{pulse} $ is positive, the natural speed $ v_0 = \dot{x} = S_0 - \gamma x$ can become negative if $S_0$ is too negative. This means that the oscillators \textit{decrease} in voltage in the absence of coupling. We avoid this unphysical regime by requiring $v_0 > 0 $ for $0 \leq x \leq1$, which leads to $\gamma_{min} \approx -0.881$.

Figure~\ref{c} shows the agreement between theory and simulation for  $\gamma < 0$, $\gamma = 0$, and $\gamma >0 $. As can be seen, $c$ declines most rapidly when $\gamma > 0$. This makes physical sense. When $\gamma > 0$, oscillators slow down as they increase in voltage, which makes them clump closer together near $x=1$. When $\gamma < 0$, the opposite happens; clusters spread further apart closer to threshold. Now suppose a $j$-cluster fires. When $\gamma > 0$ the interval $[1-j/N, 1)$ is more likely to contain oscillators than when $\gamma < 0$, thanks to the 'clumping' and 'spreading out',  which in turn makes an absorption more likely. The case of zero cavity, when $\gamma = 0$, then interpolates between these two regimes, as evidenced by Figure \ref{c}.

%%%%%%%%%%%    FIG 2      %%%%%%%
\iffalse
\begin{figure}[h!]
  \centering
\includegraphics[width=9cm]{c_gamma_point_9.png}  
\caption{(Color online) Theoretical and simulated $c(t)$ for $\gamma = 2, -0.8$. Solid lines show theoretical prediction \eqref{c_soln1}, while data points show simulated results for $N = 5 \times 10^4$ oscillators.}
\label{c}
\end{figure}
\fi

\begin{figure}[h!]
  \centering
\includegraphics[width=8.5cm]{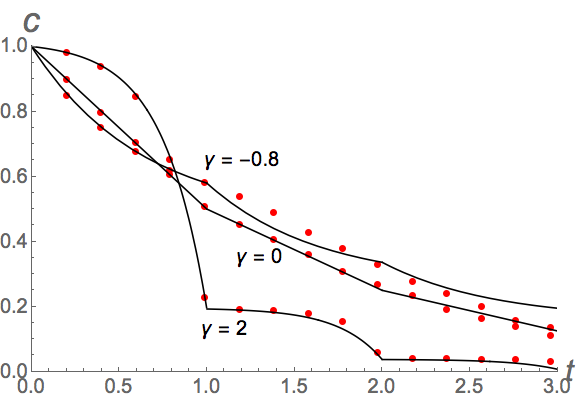}  
\caption{(Color online) Theoretical and simulated $c(t)$ for $\gamma = 2$, $\gamma = 0$, and $\gamma = -0.8$ . Solid lines show theoretical prediction \eqref{c_soln1}, while data points show simulated results for $N = 5 \times 10^4$ oscillators.}
\label{c}
\end{figure}
\

\iffalse
\begin{figure}[h!]
  \centering
\includegraphics[width=8cm]{c_diff_gammas1.png}  
\caption{(Color online) Theoretical and simulated $c(t)$ for $\gamma = 2$ and $\gamma = -0.8$ . Solid lines show theoretical prediction \eqref{c}, while data points show simulated results for $N = 5 \times 10^4$ oscillators.}
\label{c}
\end{figure}
\fi

\subsection{Individual Cluster Densities}

%%%%%%%  i = 1     %%%%%%%

How do the individual densities $c_i$ evolve? We begin with the $1$-clusters, whose density is $c_1$. They are the easiest density to analyze, since they can only decrease. There are two ways this can happen: (i) the loss of a \textit{firing} singleton, when it absorbs other clusters of any size, and (ii) the loss of \textit{absorbed} singletons, due to the firing of another cluster:

\be
\dot{c_1} = \mathcal{L}_1^{firing} + \mathcal{L}_1^{absorbed}.
\label{rate_eqn_c1}
\fe

\noindent
We begin with $\mathcal{L}_1^{firing}$. From \eqref{R_i_soln} we know singletons fire at rate $R_1 = (S_0- \gamma) e^{\gamma t} \tilde{c_1}$. During such a firing, an absorption will take place if there is at least one cluster on $[1-1/N, 1)$. This interval contains on average $N c \times 1/N = c(t) = \tilde{c} e^{\gamma t} $ clusters. Further, the probability that it contains $n$ clusters is given by the Poisson distribution: $\Pi_n = \frac{(\tilde{c} e^{\gamma t} )^n}{n!} e^{-\tilde{c} e^{\gamma t} }$. The is the mathematical statement that the clusters are distributed randomly without correlations. The probability that $[1-1/N,1)$ is occupied by at least one cluster is therefore $1 - e^{-\tilde{c} e^{\gamma t} }$. If an absorption takes place, $N_1$ decreases by $1$, since we're only considering the loss of the \textit{firing} oscillator here. The expected loss rate is then $  (S_0- \gamma) e^{\gamma t} \tilde{c_1} [ 1 \times (1 - e^{-\tilde{c} e^{\gamma t} }) + 0 \times e^{-\tilde{c} e^{\gamma t} } ]$, leading to,

\be
\mathcal{L}_1^{firing} =  (S_0- \gamma) e^{\gamma t} \tilde{c_1}(1 - e^{-\tilde{c} e^{\gamma t} }).
\fe

\noindent
To calculate $\mathcal{L}_1^{absorbed}$, imagine a $j$-cluster fires and absorbs all the singletons on the interval $[1- j/N, 1)$. As before, this interval will have on average  $N c_1(t) \times j/N = j \tilde{c_1} e^{\gamma t}$ such singletons. Multiplying this by $R_j$ and summing over $j$ then gives $\sum_j (1- \gamma) \tilde{c_j} e^{\gamma t} \times j \tilde{c_1} e^{\gamma t}$, which leads to

\be
\mathcal{L}_1^{absorbed} =   (1-\gamma) e^{2 \gamma t} \tilde{c_i}
\fe

\noindent
Substituting $\mathcal{L}_1^{firing}$ and $\mathcal{L}_1^{absorbed}$ into \eqref{rate_eqn_c1} gives,

\be
\dot{c_1} = -(S_0-\gamma) \tilde{c_1} \Big[ (1 + e^{\gamma t}) - e^{-\tilde{c} e^{\gamma t}} \Big].
\fe

\begin{figure}[h!]
  \centering
\includegraphics[width=8.5cm]{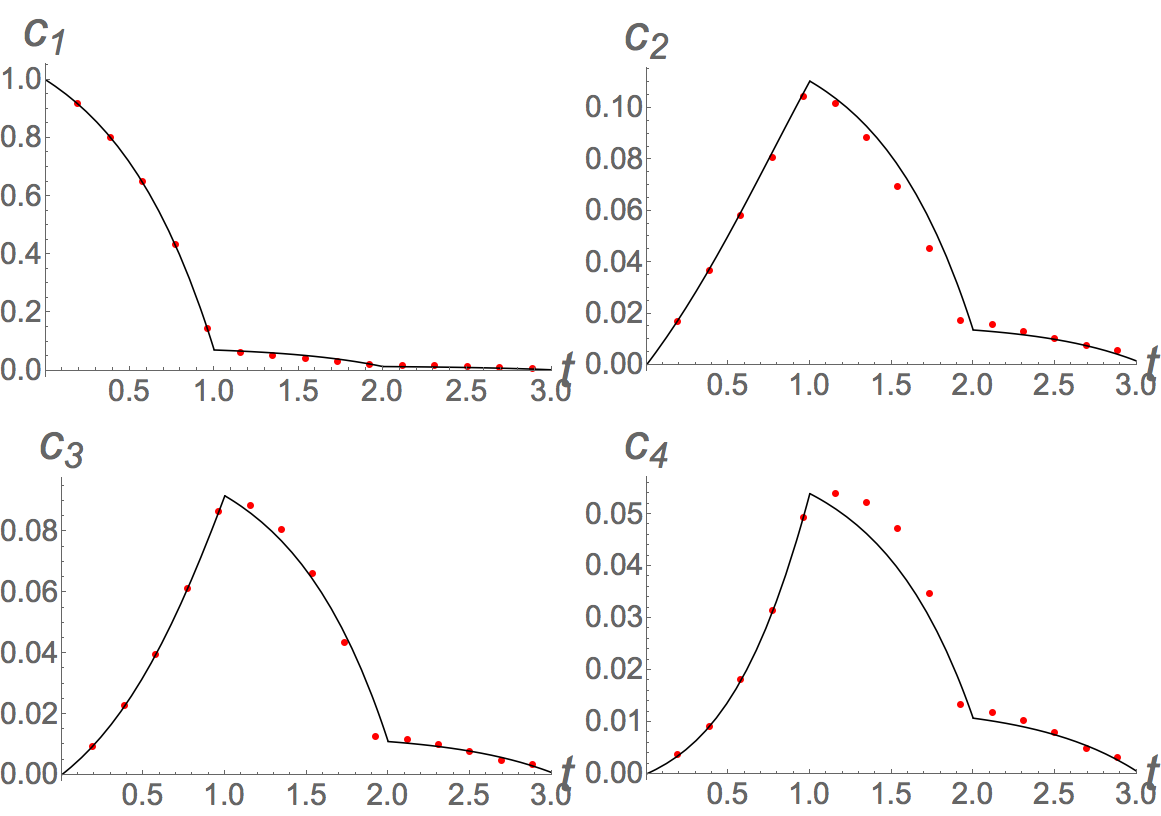}  
\caption{(Color online) Theoretical and simulated cluster densities $c_1$ though $c_4$ for $\gamma = 0.9$. Solid black lines show analytic solutions to \eqref{ci_rate_eqns}. Red data points show simulations results for $5 \times 10^4$ oscillators.}
\label{c_i_s}
\end{figure}

\noindent
This looks intimidating, but since the quantities $\tilde{c_i}$ are held constant over each period, the R.H.S. is a function of only $t$. It therefore has an analytic solution, which we show plotted in Figure~\ref{c_i_s}.

%%%%%%%  i > 1   %%%%%%%

Will larger clusters behave similarly? They differ from the singletons in that they can be created as well as absorbed, which makes them harder to calculate. Their absorption rate is easily generalized from that of the singletons:

\be
 \mathcal{L}_i^{firing} + \mathcal{L}_i^{absorbed}  = (S_0-\gamma t) \tilde{c_i} \Big[ (1 + e^{\gamma t}) - e^{-i\tilde{c} e^{\gamma t}} \Big].
 \fe

\noindent 
Their gain rate is calculated as follows. A $i$-cluster is created when a cluster of size $k<i$ fires, and absorbs the right combination of other clusters. Suppose there are $a_1$ $1$-clusters, $a_2$ $2$-clusters, $\dots$, on the interval $[1-k/N, 1)$. If $a_1 + 2a_2 + \dots + k  = i$, then an $i$-cluster will be created. Such a combination occurs with probability $\frac{(kc_1)1^a_1}{a_1 !} e^{ - kc_1} \times \frac{(kc_2)1^a_2}{a_2 !} e^{ - kc_2} \times \dots  $. Summing first over all such combinations, and then over all $k$, gives an expected rate gain of

\be
\sum_{k=1}^{i-1}(S_0- \gamma) \tilde{c_k} e^{\gamma t} e^{-k \tilde{c} e^{\gamma t}} \sum_{a_1 + 2 a_2 + \dots = i-k} \left( \prod_{p \geq 1} \frac{(k \tilde{c_p} e^{\gamma t})^{a_p}}{a_p!} \right)
\fe

\noindent
After combining the loss and gain terms, and some algebraic manipulation, we finally obtain the desired rate equation for $i$-clusters,

\be 
\begin{split}
\dot{c_i} &= - (S_0-\gamma) e^{\gamma t}(1 + e^{\gamma t})\tilde{c_i} + \\
& \sum_{k=1}^{i} (S_0- \gamma) \tilde{c_k} e^{\gamma t} e^{- k \tilde{c} e^{\gamma t}} \sum_{\sum p a_p = i-k} \left( \prod_{p \geq1} \frac{(k \tilde{c_p} e^{\gamma t})^{a_p}}{a_p !} \right) 
\end{split}
\label{ci_rate_eqns}
\fe

\noindent
This is a set of recursive equations, and so we can solve them successively. As with $c_1$, the R.H.S. is a pure function of $t$, so analytic solutions are findable. Figure~\ref{c_i_s} show theoretical prediction versus simulation results.

%%%%%%%%%%%    FIG 3      %%%%%%%

%The shape of the curves for $c_{i>2}$ look strikingly similar. This is no surprise given the recursive structure of \eqref{ci_rate_eqns}. We ask the question 

%%%%%%%%%%%   Alternate couplings     %%%%%%%

\subsection{Alternate Coupling Rules}

We thus far assumed an $i$-cluster fired a pulse of size $i/N$. We now  consider two alternatives. The first is simply the original pulse strength with a tunable strength $K$: $(Ki)/N$. The second is a fixed pulse strength of $K/N$ regardless of the size of the firing cluster. These alterations only modestly change the analysis,  so we simply list the results for $L_i, R_i, v_{pulse}$ and $c$ in the table below, where $S_0$ is given by \eqref{S_0}. For illustrative purposes we do not include a formula for $c_i$, but its calculations is straightforward.

\begin{table}[h!]
\centering
\begin{tabular}{l*{6}{|c|}r}
             & $K j/N$ & $K/N$  \\
\hline
$L_i$             &$K i \tilde{c} e^{\gamma t}$  & $ K \tilde{c} e^{\gamma t} $   \\
$R_i $   & $ (S_0-\gamma) \tilde{c}_i e^{\gamma t} $ &$ (S_0-\gamma) \tilde{c}_i e^{\gamma t} $   \\
$v_{pulse} $   & $ K (S_0-\gamma)  e^{\gamma t} $ &$ K (S_0-\gamma) e^{\gamma t} \tilde{c}  $   \\
$c(t)$          & $\frac{\tilde{c} \Big( K  \left(\gamma -S_0\right) \left(e^{2 \gamma  t}-1\right)+2\gamma  \Big)}{2 \gamma }$ & $\frac{\tilde{c} \Big( \tilde{c} K \left(\gamma -S_0\right) \left(e^{2 \gamma  t}-1\right)+2\gamma  \Big)}{2 \gamma }$  \\
\end{tabular}
\end{table}

As can be seen, there are mostly only minor differences between the two cases. We remark however that $v_{pulse}$ now depends on $\tilde{c}$ for the fixed pulse case. This makes sense physically;  since there are fewer clusters in successive periods, and the pulse per cluster is constant, the total 'current' per period will get smaller. This is in contrast to the pulse = $K j / N$ case, where there are fewer clusters per period also, but larger clusters fire larger pulses, keeping the total 'current' per period constant.  A consequence of this decrease in $v_{pulse}$ is that the periods won't be constant for pulse $= K/N$, as there are for $K j / N$. They will get longer as $v_{pulse}$ decreases from period to period.

%%%%%%%%%%%%%%%%%%%%%%%%%%%%%%%%%%%%%%%%%%%%%%%%%%%%%%%%%%%%%%%%

\section{Breakdown of Approximations} \label{approximations} 

\subsection{Uniformity Assumption}

We now discuss the approximations and assumptions we made in our analysis. The first of these was that each cluster density was distributed uniformly in voltage at the start of each period,  $\rho_j(x,t) = \tilde{N_j}$. From this, we derived equations \eqref{L_i_soln} and \eqref{R_i_soln} for $L_i$ and $R_i$, which in turn led us to our disorder parameter $c$. 

This uniformity assumption clearly cannot be satisfied for each $i$, at every $t$. For instance, consider the end of the first period. Perhaps mostly clusters of size $<5$ were formed, with only a few larger clusters of size $>10$. Then, $\rho_{j<5}(x, t = 1)$ will be approximately uniform, but $\rho_j(x, t = 1 ) $ will be more sharply peaked. So the uniformity assumption is inaccurate for large clusters, which are few in number. This explains why \eqref{c_soln1} approximates $c(t)$ well. Since $c(t) = \sum_j c_j$, we see that the sum will be dominated by those $c_i$ which are large, for which the uniformity assumption is accurate.

The fact that the uniformity assumption worsens for larger cluster sizes also means that our results for $c_i$ should get worse for larger $i$. Figure~\ref{higher_ci_s} below shows that this is indeed the case.

%%%%%%%%%%%    FIG 4      %%%%%%%
\begin{figure}[h!]
  \centering
\includegraphics[width=8cm]{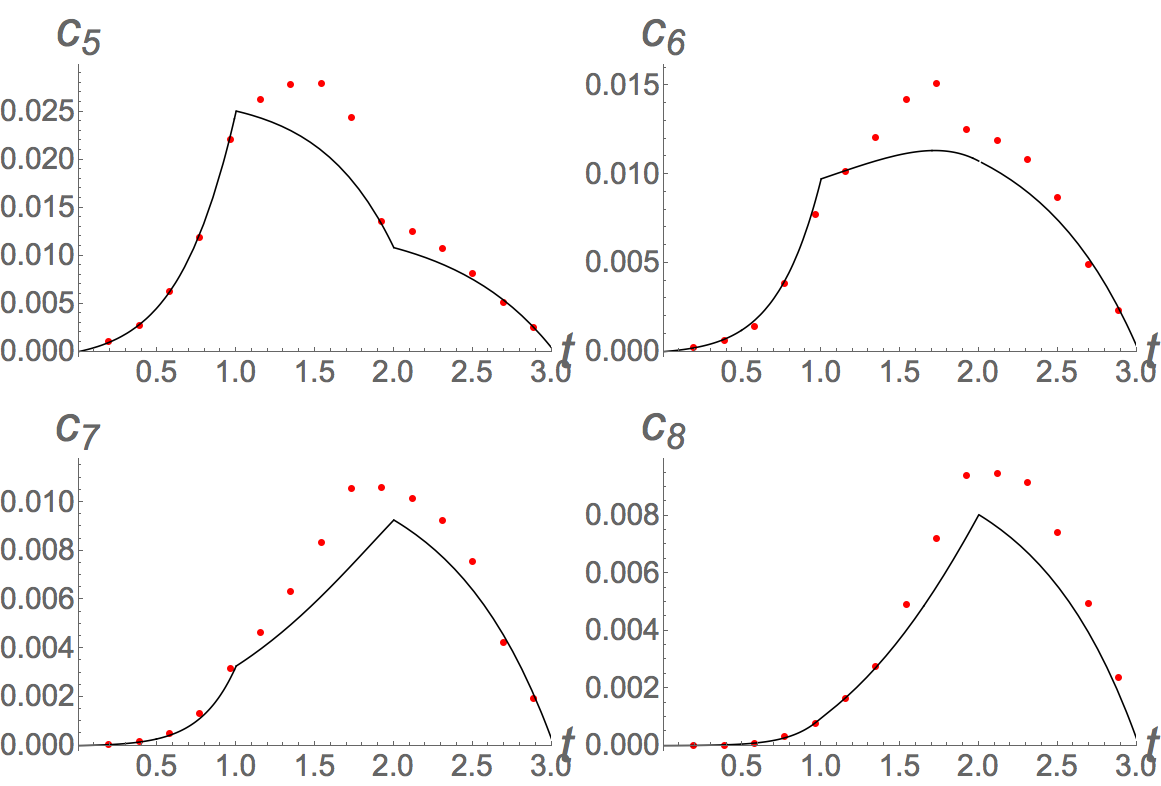}  
\caption{(Color online) Theoretical and simulated cluster densities $c_5$ though $c_8$ for $\gamma = 0.9$. Solid black lines show analytic solutions to \eqref{ci_rate_eqns}. Red data points show simulations results for $5 \times 10^4$ oscillators. As can be seen, theory and simulation start to disagree}
\label{higher_ci_s}
\end{figure}

\subsection{Final stages of process}

\begin{figure}[h!]
  \centering
\includegraphics[width=8cm]{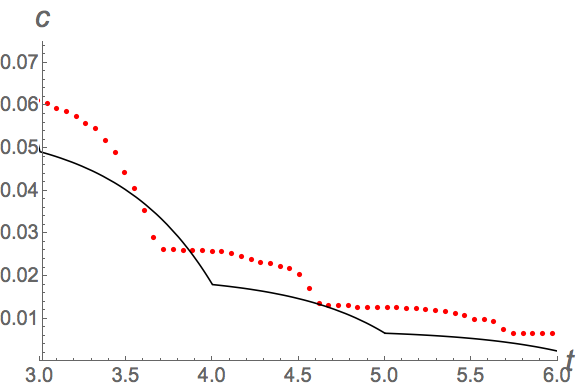}  
\caption{(Color online) Theoretical and simulated total cluster density $c(t)$ for $\gamma = 0.9$ and $t  > 3$. Solid black lines show analytic solution \eqref{c}. Red data points show simulation results for $N = 10^4$ oscillators. As can be seen, there is a significant disagreement between theory and simulation for later times, when the approximations we made in the analysis breakdown.}
\label{c_late_times}
\end{figure}

Throughout our analysis, we assumed $N_j \gg 1$. As previously discussed, this cannot be true for each $j$ at all times. This assumption is most blatantly incorrect during the final stages of the process, where there are a small number of macroscopic clusters. Our result should therefore substantially disagree with simulation for large enough $t$, as is evident in Figure~\ref{c_late_times}. \\

%%%%%%%%%%%    FIG 5      %%%%%%%

%%%%%%%%%%%%%%%%%%%%%%%%%%%%%%%%%%%%%%%%%%%%%%%%%%%%%%%%%%%%%%%%

\section{Conclusion}
We have studied the transient dynamics of pulse coupled oscillators with nonlinear charging curves. We derived approximations for the total cluster density $c(t)$ and individual cluster densities $c_i(t)$. These approximations were good up to the final stages of the process, where the assumptions made in the analysis breakdown.

A possible application of our results is in network detection. Arenas et al \cite{arenas} showed that transient clustering in the Kuramoto model can be used to approximate the underlying network structure. Could our results could be used to the same effect in networks of pulse-coupled oscillators? Local coupling would however mean that clusters could break apart as well as coalesce. One could account for this effect by including additional loss terms in our rate equations for $c$ and $c_i$, \eqref{c_rate_eqn}, \eqref{ci_rate_eqns}.

Our model has several idealizations that could be relaxed in future work. For example, delay between emission and reception of pulses, heterogeneity in oscillator speeds and pulse sizes, and non-global coupling. Another interesting modification would be to allow chain reactions, by allowing any clusters that are brought to threshold by another firing cluster, to fire themselves. 

\section{Acknowledgments}

This research was supported in part by the National Science Foundation through Grant No. DMS-1513179. We thank Steven Strogatz and Paul Krapivsky for helpful discussions.

%%%%%%%%%%%%%%%%%%%%%%%%%%%%%%%%%%%%%%%%%%%%%%%%%%%%%%%%%%%%%%%%

\section{Appendix}

We here approximate the density $\rho_j^{original}(x,t)$. For convenience, we drop the superscript 'original'.  As shown in the main body of the text, the density solves equation \eqref{cont_eqn2} below,

\begin{align}
    &\dot{\rho_j} + \partial_x(v \rho_j) = 0 \nonumber \\
    &\rho_j(x,0) = f_0(x) H(x) H(1-x) 
    \label{cont_eqn2}
  \end{align}

\noindent
where, $f_0(x) = \tilde{N_j}$ (since we are assuming a initial uniform distribution), and

\be
v(x,t) = v_0(x)  + v_{pulse}(t). 
\fe

\noindent
While we know $v_0(x) = S_0 - \gamma x$, we don't yet have a complete expression for $v_{pulse}(t)$. In the main text, we derived $v_{pulse} = \sum_j j R_j$, which using \eqref{R_i_soln} for $R_i$ gives

\be
v_{pulse}(t) = \sum_j \frac{S_0 - \gamma}{N}  j \;  \rho_j(x = 1,t)
\label{vp_step}
\fe

\noindent
This is the source of our difficulty. Our PDE for $\rho_j(x,t)$ depends on $v_{pulse}$, which in turn depends on the voltage density for \textit{every other} cluster size $\rho_k(x,t)$. To overcome this difficulty, we use a technique similar to the 'leap-frog' or 'split' method used in certain numerical schemes. This involves making a series of recursive approximations for $v_{pulse}$ and $\rho_j$:

\begin{align}
 v_{pulse} &=  \Big( v_{pulse}^{(0)},  v_{pulse}^{(1)},  v_{pulse}^{(2)}, \dots \Big) \\
 \rho_j &= \Big( \rho_j^{(0)},  \rho_j^{(1)},  \rho_j^{(2)}, \dots \Big)
\end{align}

\noindent
Graphically, our scheme is given by the following, where we have placed the labels of equations used to make the approximations over the arrows.

\be
v_{pulse}^{(0)} \xrightarrow{\eqref{cont_eqn2}} \rho_j^{(0)} \xrightarrow{\eqref{vp_step}} v_{pulse}^{(1)} \xrightarrow{\eqref{cont_eqn2}}  \rho_j^{(1)} \xrightarrow{\eqref{vp_step}} + \dots
\fe

\noindent
We begin by setting $v_{pulse}^{(0)} = 0$. The speed is then,

\be
v(x,t)^{(0)} =  v_0(x) + 0  = S_0 - \gamma x.
\fe

\noindent
We plug this into \eqref{cont_eqn2} and solve for resulting PDE for $\rho_j^{(0)}(x,t)$. This has solution,

\begin{equation}
\rho_j^{(0)}(x,t) = e^{\gamma t} \tilde{N_j} H \Big( \Gamma_0(x,t) \Big) H \Big( 1 -\Gamma_0(x,t) \Big). 
 \label{rho0}
\end{equation}

\noindent
with $ \Gamma_0(x,t) = \gamma^{-1} [ S_0 - (S_0 - \gamma x) e^{\gamma t} ] $. We then use $\rho_j^{(0)}$ to find $v_{pulse}^{(1)}$ using \eqref{vp_step}, which gives

\be
v_{pulse}^{(1)} = (S_0 - \gamma ) e^{\gamma t} \tilde{c}.
\label{v1_pulse}
\fe

\noindent
This completes the first step of our scheme. We then repeat the process to find $\rho_j^{(1)}$ and $v_{pulse}^{(2)}$. We use $v_{pulse}^{(1)}$ to update the speed,

\be
\begin{split}
v(x,t)^{(1)} &=  v_0(x) + v_{pulse}^{(1)}  \\
                 &= (S_0 - \gamma x) + (S_0 - \gamma ) e^{\gamma t} \tilde{c}.
\end{split}
\fe

\noindent
and then plug this into \eqref{cont_eqn2} to obtain a PDE for $\rho_j^{(1)}$, which we solve to get,

\begin{equation}
 \rho_j^{(1)}(x,t) = e^{\gamma t}  \tilde{N_j} H \Big( \Gamma_1(x,t) \Big) H \Big( 1 -\Gamma_1(x,t) \Big) 
 \label{rho1}
\end{equation}

\noindent
where \scalebox{0.95}{$ \Gamma_1(x,t) = \Big[ \frac{3S_0-\gamma}{2 \gamma } +\frac{e^{\gamma  t}}{2 \gamma } \left(2 \gamma  x-2 S_0\right) + \frac{e^{2 \gamma  t}}{2 \gamma}  \left(\gamma -S_0\right) \Big] $}. 

Looking at \eqref{rho0} and \eqref{rho1}, we see that $\rho_j^{(0)}$ and $\rho_j^{(1)}$ have the same functional form. They only differ in the arguments of the Heaviside function:  $\Gamma_0(x,t) \neq \Gamma_1(x,t)$. This remarkable similarity between $\rho_j^{(0)}$ and $\rho_j^{(1)}$ has an important consequence: it 'closes' our approximation scheme. We see this by substituting $\rho_j^{(1)}$ into \eqref{vp_step} to find,

\be
v_{pulse}^{(2)}  = (S_0 - \gamma ) e^{\gamma t} \tilde{c} = v_{pulse}^{(1)},
\fe

\noindent
which implies that $\rho_j^{(2)} = \rho_j^{(1)}$, which in turn implies our scheme terminates at $(v_{pulse}, \rho_j) = (v_{pulse}^{(2)}, \rho_j^{(1)})$. Our final approximations for $v_{pulse}$ and $\rho_j(x,t)$ are then,

\begin{align}
& v_{pulse}(t) \approx (S_0 - \gamma ) e^{\gamma t} \tilde{c}  \\
& \rho_j(x,t) \approx e^{\gamma t}  \tilde{N_j} H \Big( \Gamma(x,t) \Big) H \Big( 1 -\Gamma(x,t) \Big)  
\end{align}

\noindent
with \scalebox{0.95}{$ \Gamma(x,t) = \Big[ \frac{3S_0-\gamma}{2 \gamma } +\frac{e^{\gamma  t}}{2 \gamma } \left(2 \gamma  x-2 S_0\right) + \frac{e^{2 \gamma  t}}{2 \gamma}  \left(\gamma -S_0\right) \Big] $}. \\

This concludes our analysis. We state bluntly that our approach is not rigorously justified. Its legitimacy is  supported only by the agreement between our analytic results and numerical simulation. We hope future work will elucidate the cause of its efficacy.

\end{document}